\documentclass[sigconf]{acmart}
\usepackage{threeparttable}
\usepackage{color}
\usepackage{booktabs} 
\usepackage{amsmath}
\usepackage{amsfonts}
\usepackage{comment}
\usepackage{footmisc}
\usepackage{mathrsfs}
\usepackage{graphicx}
\usepackage{subfigure}
\usepackage{multirow}
\usepackage{algorithm}
\usepackage{algorithmic}
\usepackage{multicol}  
\usepackage{enumitem}
\usepackage{array}
\usepackage{float}
\usepackage{hyperref}

\usepackage{soul, color, xcolor,xspace}
\newcommand{\name}{LinRec\xspace}

\copyrightyear{2023}
\acmYear{2023}
\setcopyright{acmlicensed}\acmConference[SIGIR '23]{Proceedings of the 46th International ACM SIGIR Conference on Research and Development in Information Retrieval}{July 23--27, 2023}{Taipei, Taiwan}
\acmBooktitle{Proceedings of the 46th International ACM SIGIR Conference on Research and Development in Information Retrieval (SIGIR '23), July 23--27, 2023, Taipei, Taiwan}
\acmPrice{15.00}
\acmDOI{10.1145/3539618.3591717}
\acmISBN{978-1-4503-9408-6/23/07}

\begin{document}

\title{LinRec: Linear Attention Mechanism for Long-term Sequential Recommender Systems}
\author{Langming Liu}
\affiliation{%
  \institution{City University of Hong Kong}
  \country{}
}

\author{Liu Cai}
\affiliation{
  \institution{Ant Group}
  \country{}
}

\author{Chi Zhang}
\affiliation{%
  \institution{Harbin Engineering University}
  \country{}
}
\author{Xiangyu Zhao}
\authornote{Xiangyu Zhao is corresponding author. xianzhao@cityu.edu.hk}
\affiliation{%
  \institution{City University of Hong Kong}
  \country{}
}

\author{Jingtong Gao}
\affiliation{%
  \institution{City University of Hong Kong}
  \country{}
}
\author{Wanyu Wang}
\affiliation{%
  \institution{City University of Hong Kong}
  \country{}
}
\author{Yifu Lv}
\affiliation{
  \institution{Ant Group}
  \country{}
}
\author{Wenqi Fan}
\affiliation{%
  \institution{The Hong Kong Polytechnic University}
  \country{}
}
\author{Yiqi Wang}
\affiliation{%
  \institution{National University of Defense Technology}
  \country{}
}
\author{Ming He}
\affiliation{%
  \institution{AI Lab, Lenovo Research}
  \country{}
}
\author{Zitao Liu}
\affiliation{%
  \institution{Guangdong Institute of Smart Education, Jinan University}
  \country{}
}
\author{Qing Li}
\affiliation{%
  \institution{The Hong Kong Polytechnic University}
  \country{}
}

\renewcommand{\shortauthors}{Langming Liu, et al.}
\begin{abstract}

Transformer models have achieved remarkable success in sequential recommender systems (SRSs). 
However, computing the attention matrix in traditional dot-product attention mechanisms results in a quadratic complexity with sequence lengths, leading to high computational costs for long-term sequential recommendation.
Motivated by the above observation, we propose a novel L2-Normalized \textbf{\underline{Lin}}ear Attention for the Transformer-based Sequential \textbf{\underline{Rec}}ommender Systems (LinRec), which theoretically improves efficiency while preserving the learning capabilities of the traditional dot-product attention.
Specifically, by thoroughly examining the equivalence conditions of efficient attention mechanisms, we show that LinRec possesses linear complexity while preserving the property of attention mechanisms. In addition, we reveal its latent efficiency properties by interpreting the proposed LinRec mechanism through a statistical lens. Extensive experiments are conducted based on two public benchmark datasets, demonstrating that the combination of LinRec and Transformer models achieves comparable or even superior performance than state-of-the-art Transformer-based SRS models while significantly improving time and memory efficiency. The implementation code is available online at \url{https://github.com/Applied-Machine-Learning-Lab/LinRec}.
\end{abstract}


\maketitle

\section{Introduction}
In recent years, sequential recommender systems (SRSs) have become an increasingly popular and widely-applied technology~\cite{wang2015learning,he2018translation,wang2019sequential,huang2018improving,tang2019towards,chen2022intent,xie2022contrastive,qiu2022contrastive,zhang2021causerec}, with applications in various \textcolor{black}{practical scenarios} such as social media, e-commerce, and online movie platforms
\cite{liu2023multi,ge2021towards,fan2021attacking,liu2023exploration,zhao2022adaptive,zhao2021dear,zhao2019deep,zhao2018deep,ren2022variational}. 
In practice, users' historical interaction sequences are typically long-term, which contain valuable yet unequal information (i.e., different interactions' importance), including more and less recent interactions, for revealing users' actual preferences~\cite{zhao2022mae4rec,zhang2020deep,zou2020neural,zhang2023denoising,yang2023generating,zhao2020whole,zhao2018recommendations,zhao2017deep,zhao2020jointly,zhang2023variational,zhang2022hierarchical,zheng2022ddr}. Therefore, identifying important interactions while not losing valuable information from a sequence~\cite{wang2019sequential, chen2018sequential}, thus learning better sequence representation for making next-item recommendations, leads to the problem of \textit{long-term sequential recommendation}.

Towards this purpose, the Transformer architecture~\cite{vaswani2017attention} has gained significant attention since its capabilities for learning informative long-term sequential patterns among historical user-item interactions.
The core component of Transformer-based models is the dot-product attention mechanism~\cite{vaswani2017attention}, which computes the corresponding attention matrix for distinguishing items' importance by a dot-product operation between the \textit{query} and \textit{key} matrices (Query and Key for short), thus learning sequence representations.
For example, BERT4Rec~\cite{sun2019bert4rec} uses multi-head self-attention and simultaneously calculates the attention score of all positions.
FDSA~\cite{zhang2019feature} introduces multiple attention blocks to depict the potential features. SASRec~\cite{kang2018self} controls the predictions based on only a small amount of actions by adopting an attention mechanism. However, a significant limitation of these models arises when dealing with long-term sequences, where the sequence length $N$ is far greater than the item embedding size $d$ ($N\gg d$). Due to the dot-product operation in the attention layer, \textcolor{black}{the computational and memory complexity of transformers are $\mathcal{O}(N^2)$}, resulting in the complexity could be dramatically increased with the problem scale (i.e., sequence lengths) for long-term sequential recommendation~\cite{li2023automlp,li2022mlp4rec,liang2023mmmlp,zhao2023user}.

In view of the above limitations, several approaches have been proposed to address the high computational costs of Transformer models, including Fixed Patterns (FP), Combination of Patterns (CP), Learnable Patterns (LP), Neural Memory, Low-Rank Methods, Kernels, and Recurrence~\cite{tay2022efficient}. Some of these categories, such as FP, LP, and low-Rank, are theoretically less complex than the standard dot-product attention and have been shown to effectively reduce memory and time costs in practice~\cite{qiu2019blockwise, parmar2018image,wang2020cluster,kitaev2020reformer,child2019generating,
beltagy2020longformer,wang2020linformer,katharopoulos2020transformers}. Notably, there are several efficient transformer models with linear complexity, such as Linformer~\cite{wang2020linformer}, Linear Transformers~\cite{katharopoulos2020transformers}, and Big Bird~\cite{zaheer2020big}. However, these methods may not be well-suited for SRSs, particularly when dealing with long-term sequences, as they often introduce additional steps to improve efficiency while sacrificing accuracy and stability. For instance, to reduce the rank of the attention matrix, Linformer~\cite{wang2020linformer} introduces the projection for both Query and Key, which results in additional complexity and impairs accuracy, jeopardizing recommendation performance.

In this paper, we propose an efficient attention mechanism to address the issue of high complexity transformers with linear complexity, called L2-normalized \textbf{Lin}ear attention for long-term sequential \textbf{Rec}ommender systems (\name). Our proposed method aims to create a method for the long-term sequential recommender that not only retains the advantages of attention (such as high accuracy) but also significantly reduces the computational complexity. Additionally, \name can be conveniently transplanted to any transformer for sequential recommendation systems, providing flexibility and compatibility. Moreover, \name does not introduce additional steps, but directly enhances the attention layer to improve efficiency, preserving both effectiveness and stability.
To be more specific, the proposed \name mechanism involves three key modifications compared to standard attention: (\textit{i}) changing the dot-product order of the attention mechanism, (\textit{ii}) using row-wise and column-wise normalization methods for Query ($\boldsymbol{Q}$) and Key ($\boldsymbol{K}$) respectively, and (\textit{iii}) adding an activation layer to $\boldsymbol{Q}$ and $\boldsymbol{K}$. These modifications enable \name to reduce the complexity from $\mathcal{O}(N^2)$ to $\mathcal{O}(N)$, while still preserving the attention property and providing sparsity.

The major contributions to our work are four-fold:
\vspace{-2mm}
\begin{itemize}[leftmargin=*]
\item We develop a novel L2 normalized linear attention (\name) for long-term sequential recommendations, which reduces the complexity of the attention mechanism from $\mathcal{O}(N^2)$ to $\mathcal{O}(N)$, while preserving the high accuracy of the attention mechanism; 

\item We theoretically analyze the proposed \name mechanism, including its effectiveness and efficiency, justifying the correctness of our design choice. Moreover, we explain and factorize the attention operation from a statistical perspective, demonstrating the inherent relation of efficient Transformer with probabilities;

\item The proposed \name mechanism is generally applicable to most transformer models for SRSs, as it can be easily incorporated into existing transformers by replacing the standard dot-product attention, providing flexibility and compatibility;

\item Empirical evaluations are conducted on two public benchmark datasets (ML-1m and Gowalla), which demonstrates that \name possesses competitive or superior performance than representative Transformer-based SRS models, while greatly reducing computational and memory costs.
\end{itemize}

\section{Preliminary}
\label{sec:preliminary}
In this section, we briefly introduce and discuss the widely used dot-product attention mechanism in Transformer-based models and the long-term sequential recommendation.

\subsection{Dot-Product Attention}
The critical part of transformers is the attention layer. The central concept underlying the attention mechanism is that each element in the sequence should learn to collect information from other tokens. Below is a standard dot-product attention. First, we define $N$ as sequence length and $d$ as hidden size. The standard attention mechanism can be written as:
\[\boldsymbol{A}=\mathrm{softmax}(\frac{\boldsymbol{Q}\boldsymbol{K}^\mathrm{T}}{\sqrt{d}})\boldsymbol{V},\]
\[where \quad \boldsymbol{Q}=\boldsymbol{X}\boldsymbol{W}_Q, \boldsymbol{K}=\boldsymbol{X}\boldsymbol{W}_K, \boldsymbol{V}=\boldsymbol{X}\boldsymbol{W}_V,\]
where $\boldsymbol{X}\in \mathbb{R}^{N\times d}$ is the input sequence matrix, $\boldsymbol{W}_Q, \boldsymbol{W}_K, \boldsymbol{W}_V$ are weight matrices for the projections that are learned from the training process, $\boldsymbol{Q}, \boldsymbol{K}, \boldsymbol{V}\in \mathbb{R}^{N\times d}$ are Query, Key and Value matrices which represent all $N$ positions' queries, keys, and values, $\mathrm{softmax}(\cdot)$ is row-wise softmax.
$\boldsymbol{Q}\boldsymbol{K}^\mathrm{T}$ is usually divided by $\sqrt{d}$ for scaling, and $\boldsymbol{A}$ is the output of attention mechanism.
\begin{figure}[t]
  \centering
  \includegraphics[width=0.66\linewidth]{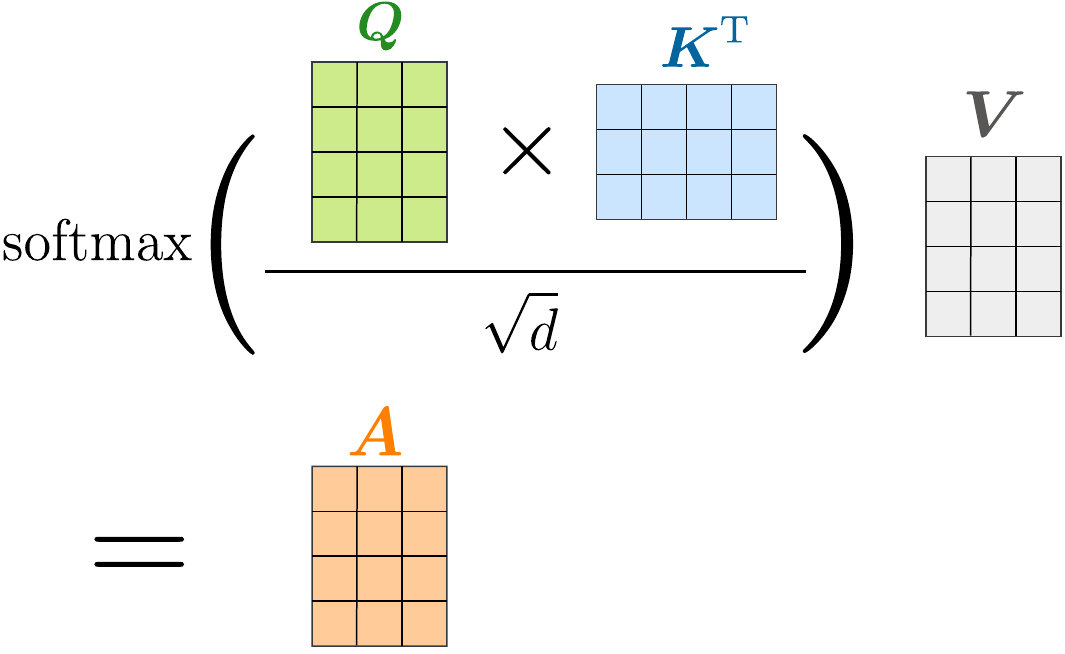}
  \vspace{-3mm}
  \caption{The standard process of Dot-Product Attention.}
  \label{fig:Figure_1}
  \vspace{-6mm}
\end{figure}

Figure~\ref{fig:Figure_1} shows that the standard attention mainly includes two dot-product operations, where Cross represents matrix multiplication. One of the highlights of the attention mechanism is that after the dot-product operation of Query, Key, and Value, the dimension of the output matrix will not vary. The dimensional invariance can convey the attention information to the sequence smoothly. 

\noindent\textbf{Advantages and Disadvantages.}
The attention matrix $\boldsymbol{Q}\boldsymbol{K}^\mathrm{T}$ is vital for the attention mechanism. Firstly, this $N\times N$ attention matrix provides scores between elements in the sequence, which helps learn the latent self-correlation in the sequence. In addition, using a matrix that consists of all sequence information reduces sequence operations and alleviates the gradient vanishing problem compared to RNN.
\begin{figure*}[t]
\vspace{-2mm}
  \centering
  \includegraphics[width=0.9\linewidth]{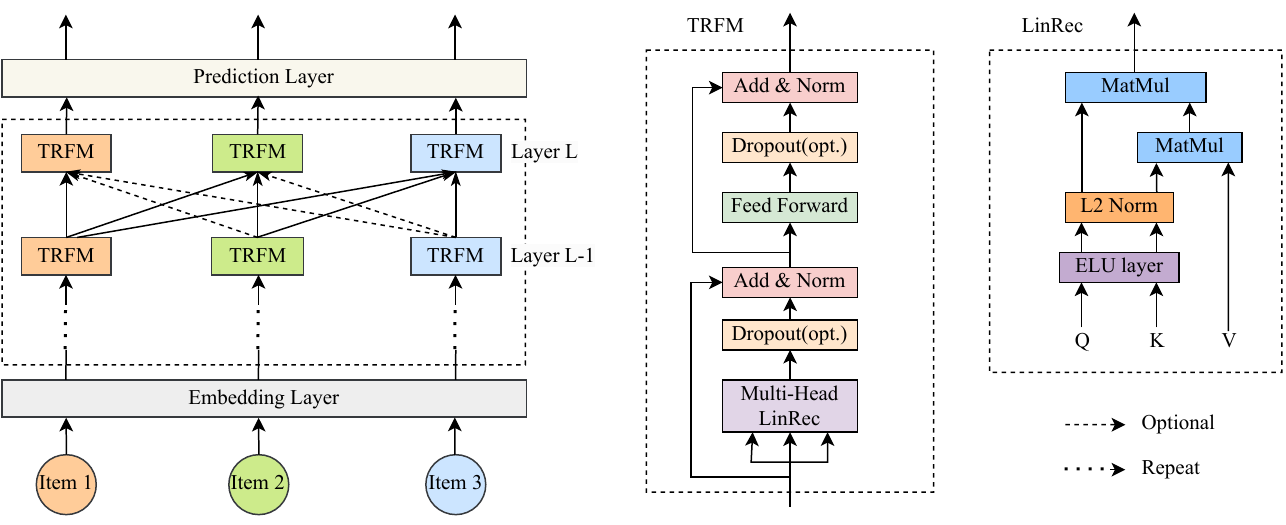}
  \vspace{-2mm}
  \caption{The overall architecture of the proposed model.}
  \label{fig:Architecture}
  \vspace{-4mm}
\end{figure*}
However, the disadvantage of attention is often about the complexity problem. The main complexity is generated by the dot-product operation of attention matrix, and the complexity is $\mathcal{O}(N^2 d)$, quadratic to the sequence length $N$. The quadratic complexity will bring great memory and time costs when $N$ is big enough. And the complexity is simplified to $\mathcal{O}(N^2)$ when $N\gg d$. Therefore, utilizing a standard attention mechanism when dealing with long-term sequences is impractical. Besides, softmax layer aggregates attention scores on a few positions, which is not conducive to mastering more information from long-term sequences.

\vspace{-2mm}
\subsection{\textbf{Long-term Sequential Recommender Systems}} 


Long-term sequential recommender systems, i.e., Long-term SRSs, are defined mathematically as systems that take into account a user's historical interactions over an extended period rather than just their recent interactions. The definition of ``long-term'' can vary depending on the specific application and context. In general, long-term SRSs are characterized by the number of historical interactions that are considered when generating recommendations, denoted as $N$. A common way to define the threshold between short-term and long-term SRSs is based on the ratio $N/d$, where $d$ is the dimension of the user or item embeddings, typically 64 to 128. In practice, a ratio of $N/d$ greater than $1.5$ is considered long-term, meaning that the number of historical interactions is at least $1.5$ times the dimension of the embeddings. For example, if $d = 64$, a long-term SRS would consider at least $100$ historical interactions.

\section{Methodology}

Given a set of user $\mathcal{U}=\{u_1,u_2,\cdots,u_{\vert\mathcal{U}\vert}\}$ and a set of item $\mathcal{V}=\{v_1,v_2,\cdots,v_{\vert\mathcal{V}\vert}\}$. Suppose that $u_i$'s historical interaction sequence is $s_i=[v_{1}^i,\cdots,v_{t}^i,\cdots,v_{n_i}^i]$, where $v_{t}^i$ is the $t$-th item interacted by user $u_i$, and $n_i$ represents the sequence' length. For each user $u_i$, \name takes $u_i$'s interaction sequence $s_i$ as input, and outputs the top-$k$ items from $\mathcal{V}$ that are most likely to be interacted with in the next time step. As illustrated in Figure~\ref{fig:Architecture}, our proposed architecture consists of three major components, including the embedding layer to generate item embeddings, the transformer layer to learn sequence representation, and the final prediction layer. In particular, to improve efficiency and applicability for Long-term SRSs, we propose a novel attention mechanism in the transformer layer. In the following, we detail each component.

\subsection{Embedding Layer}
To jointly train \name with other Transformer-based backbones, we build two parameter matrices (i.e., $\boldsymbol{E}^s$ and $\boldsymbol{E}^p$) as embedding look-up tables for items' embedding initialization in a sequence via
\begin{equation}
    \boldsymbol{E}=[\boldsymbol{E}_1^s+\boldsymbol{E}_1^p,\boldsymbol{E}_2^s+\boldsymbol{E}_2^p,\cdots,\boldsymbol{E}_N^s+\boldsymbol{E}_N^p]^{\mathrm{T}},
\end{equation} 
where $\boldsymbol{E}^s$, $\boldsymbol{E}^p\in\mathbb{R}^{N\times d}$ are trainable matrices mapping items' IDs and position in sequence into $d$-dimension dense vectors.

\subsection{Transformer Layer}
With the above embedding layer, we are ready to integrate \name into existing Transformer-based recommenders to learn sequence representations. As mentioned in Sec.~\ref{sec:preliminary}, traditional dot-product attention methods are inherently sub-optimal since they typically need high computational costs (i.e., $\mathcal{O}(N^2 d)$) to calculate the attention matrix. To preserve dot-product attention's learning capabilities while reducing its computational cost, we consequently propose the \name mechanism for long-term SRSs. Specifically, we first analyze dot-product attentions' properties to derive equivalence conditions that inspire our method design. We then propose an efficient L2 Normalization method to modify the mapping (i.e., Softmax) of the dot-product, thus switching the calculation order (i.e., compute $\boldsymbol{K}^{\mathrm{T}}\boldsymbol{V}$ first) to reduce computational complexity. Moreover, we theoretically analyze the proposed \name, justifying its correctness and computational efficiency. 
Besides, we interpret \name's superiority from a statistical perspective attached in Appendix~\ref{appendix}.

\subsubsection{\textbf{Equivalence Conditions}}
Generally, dot-product attention mechanisms~\cite{vaswani2017attention} calculate the corresponding attention matrix could be formulated as follows:
\begin{equation}
\boldsymbol{A}=\rho(\boldsymbol{Q}\boldsymbol{K}^\mathrm{T})V \:;\: 
\boldsymbol{B}=\rho(\boldsymbol{Q}\boldsymbol{K}^\mathrm{T}),
\label{eq:dot-product attention}
\end{equation}
where $\rho(\cdot)$ means the scaling and row-wise Softmax operators. Accordingly, the attention matrix $\boldsymbol{B}$ satisfies the following two properties: (1) \textit{Normalized Property}. Each row of $\boldsymbol{B}$'s elements sums up to $1$. (2) \textit{Non-Negative Property}. Each element of $\boldsymbol{B}$ is non-negative. Each row of $\boldsymbol{B}$'s elements is constrained into a range as $[0,1]$ to easily distinguish the importance of different positions (i.e., element value is proportional to the corresponding item importance) in a sequence and ensure numerical stability.

In practice, the computational complexity of computing $\boldsymbol{B}$ first could be dramatically increased with the problem scale (i.e., sequence length) ~\cite{qiu2019blockwise,liu2018generating}, which harms dot-product attentions' applicability for long-term sequences. Besides, $\rho(\cdot)$ makes $\rho(\boldsymbol{Q}\boldsymbol{K}^\mathbf{T})\boldsymbol{V}\neq\rho(\boldsymbol{Q})\rho(\boldsymbol{K}^\mathbf{T})\boldsymbol{V}$, which renders a non-trivial technical challenge (i.e., compute $\boldsymbol{B}$ last) for simplifying Eq.~(\ref{eq:dot-product attention})'s complexity. To tackle such a challenge, we could devise another mapping instead of $\rho(\cdot)$ to decompose Eq.~(\ref{eq:dot-product attention}). Thus we could compute $\boldsymbol{K}^\mathrm{T}\boldsymbol{V}$ first to reduce computation complexity without jeopardizing attention's learning capabilities.
Formally, we define such a mapping as follows:
\begin{equation}\boldsymbol{A}'=\rho_1(\boldsymbol{Q})\rho_2(\boldsymbol{K})^\mathrm{T} \boldsymbol{V}
\: ; \:
\boldsymbol{B}'=\rho_1(\boldsymbol{Q})\rho_2(\boldsymbol{K})^\mathrm{T},
\label{eq:new attention}
\end{equation}
where $\rho_1(\cdot)$ and $\rho_2(\cdot)$ are different mappings from $\mathbb{R}^{N\times d}$ to $\mathbb{R}^{N\times d}$, decomposing the original derivation (i.e., $\rho(\boldsymbol{Q}\boldsymbol{K}^\mathbf{T})$) into two parts (i.e., $\rho_1(\boldsymbol{Q})\rho_2(\boldsymbol{K})^\mathrm{T}$) for subsequent calculation.
Therefore, we could specify the $i$-th row of $\boldsymbol{B}'$ via
\begin{equation}
\boldsymbol{B}'_i=\Big(\sum_{j=1}^d \boldsymbol{Q}^\rho_{ij}\boldsymbol{K}^\rho_{1j},\sum_{j=1}^d \boldsymbol{Q}^\rho_{ij}\boldsymbol{K}^\rho_{2j},\cdots, \sum_{j=1}^d \boldsymbol{Q}^\rho_{ij}\boldsymbol{K}^\rho_{Nj}\Big),
\end{equation}
where $\boldsymbol{Q}^\rho=\rho_1(\boldsymbol{Q}),\boldsymbol{K}^\rho=\rho_2(\boldsymbol{K})$. 
Ideally, to satisfy the above requirements, we could derive two equivalence conditions as follows:
\begin{equation*}\label{eq:eqcondition_1}
\;\,\textbf{Condition (1).} \sum_{m=1}^N\sum_{j=1}^d \boldsymbol{Q}^\rho_{ij}\boldsymbol{K}^\rho_{mj}\le 1, \qquad\forall i=1,2,\cdots,N;
\end{equation*}
\begin{equation*}\label{eq:eqcondition_2}
\textbf{Condition (2).} \sum_{j=1}^d \boldsymbol{Q}^\rho_{ij}\boldsymbol{K}^\rho_{mj}\ge 0, \qquad\forall i,m=1,2,\cdots,N.
\end{equation*}
Ultimately, an appropriate design of the above mappings could enable us to compute $(\boldsymbol{K}^\rho)^\mathbf{T}\boldsymbol{V}$ first and achieve $\boldsymbol{O}(N d^2)$ complexity. This is because $\boldsymbol{K}$ and $\boldsymbol{V}$ are both $N \times d$ matrices. Besides, omitting $d$ ($N >> d$) and utilizing linear complexity mappings (i.e., $\rho_1(\cdot)$ and $\rho_2(\cdot)$) for long-term SRSs, we could achieve approximated linear complexity (i.e., $\boldsymbol{O}(N)$). However, linear complexity mappings are hard to satisfy the above conditions strictly. Toward this end, we further relax the above conditions and design the two mappings for approximating dot-product attention's learning capabilities (i.e., generate a comparable attention matrix for identifying items' importance) while significantly reducing complexity. Functionally speaking, such a proper mapping design should have the following requirements: (1) $\rho_1(\cdot)$ and $\rho_1(\cdot)$ do not introduce additional computational costs, and (2) they have to maintain attention's identifying capabilities and the numerical stability, thus learning better sequence representations to make recommendations for long-term SRSs efficiently and effectively.

\subsubsection{\textbf{L2 Normalized Linear Attention Mechanism (LinRec)}}
\label{sec:LinRec}
According to the above conditions, we further derive an equivalent condition to provide theoretically designing insights. Mathematically, Condition (1) is equivalent to the following formulation:
\begin{equation}
\sum_{j=1}^d \boldsymbol{Q}^\rho_{ij}\sum_{m=1}^N \boldsymbol{K}^\rho_{mj}\le 1,
\end{equation}
and its sufficient condition is straightforward:
\begin{equation*} \label{eq:sufficientcondition}
\textbf{Condition (3).}\sum_{j=1}^d \boldsymbol{Q}^\rho_{ij}\le 1, \quad\sum_{m=1}^N \boldsymbol{K}^\rho_{mj}\le 1.
\end{equation*}
Therefore, it is formally equivalent to L$p$ Normalization families, which constrains the summations in a small range. The L$p$ Normalization families represent a trade-off between the performance and computational cost of serving as mapping functions: a small value of $p$ could introduce fewer computational costs with relatively inaccuracy approximation. 
Specifically, we leverage row- and column-wise L2 Normalization methods to perform the two mappings. This is because L2 Normalization is more efficient than L1 Normalization and more effective than others (e.g., $p=\infty$). 
For $i$-th row of $\boldsymbol{Q}$ (i.e., $\boldsymbol{Q}_i=(\boldsymbol{Q}_{i1},\cdots,\boldsymbol{Q}_{id})$) and the $j$-th column of $\boldsymbol{K}$ (i.e., $\boldsymbol{K}_j=(\boldsymbol{K}_{1j},\cdots,\boldsymbol{K}_{Nj})^\mathrm{T}$), we have:
\begin{equation}
\begin{aligned}
&\rho_1(\boldsymbol{Q}_i)=\frac{\boldsymbol{Q}_i}{\sqrt{d}\Vert \boldsymbol{Q}_i\Vert_2}=\frac{1}{\sqrt{d}\Vert \boldsymbol{Q}_i\Vert_2}(\boldsymbol{Q}_{i1},\cdots,\boldsymbol{Q}_{id}),\\
&\rho_2(\boldsymbol{K}_j)=\frac{\boldsymbol{K}_j}{\sqrt{N}\Vert \boldsymbol{K}_j\Vert_2}=\frac{1}{\sqrt{N}\Vert \boldsymbol{K}_j\Vert_2}(\boldsymbol{K}_{1j},\cdots,\boldsymbol{K}_{Nj})^\mathrm{T},
\end{aligned}
\end{equation}
where $\Vert\cdot\Vert_2$ represents the L2 Normalization.

\noindent \textbf{Proof}: by Cauchy-Schwarz inequality ~\cite{wu2009various}, we obtain:
\begin{align*}
(\boldsymbol{Q}_{i1}+\cdots+\boldsymbol{Q}_{id})^2&=\Big(\sqrt{1\cdot \boldsymbol{Q}_{i1}^2}+\cdots+\sqrt{1\cdot \boldsymbol{Q}_{id}^2}\Big)^2\\
&\le(1+\cdots+1)(\boldsymbol{Q}_{i1}^2+\cdots+\boldsymbol{Q}_{id}^2)\\
&= d(\boldsymbol{Q}_{i1}^2+\cdots+\boldsymbol{Q}_{id}^2)\\
&=d\Vert \boldsymbol{Q}_i\Vert_2^2,
\end{align*}
then we can have:
\begin{equation}\boldsymbol{Q}_{i1}+\cdots+\boldsymbol{Q}_{id}\le\sqrt{d}\Vert \boldsymbol{Q}_i\Vert_2.\end{equation}
Therefore, $\rho_1(\cdot)$ and $\rho_2(\cdot)$ satisfy Condition (3), ensuring the normalized property of the generated attention matrix $\boldsymbol{B}'$. 

However, such normalization methods do not necessarily satisfy the non-negative condition. We consequently incorporate the Rectified Linear Unit (ReLU)~\cite{glorot2011deep} function families to eliminate negative values' influence. Although utilizing the standard ReLU function could strictly satisfy non-negativity, such an activated function may meet zero-gradient error~\cite{hendrycks2016gaussian,xu2015empirical,klambauer2017self}, which leads to an unstable training process. Besides, the ReLU function could meet inevitable information loss issues by fully setting negative elements to zero since each item should contribute to representation learning in a sequence. Therefore, we use a variation of the ReLU families, ELU function~\cite{clevert2015fast}, to perform activation and control negative values. Specifically, we formulate such a function via
\begin{equation}
\mathrm{elu}(x):=\left\{
\begin{aligned}
&x,&\text{if }&x\ge 0,\\
&e^x-1,\quad&\text{if }&x<0.
\end{aligned}\right.
\end{equation}
Accordingly, utilizing the $\mathrm{elu}(\cdot,\alpha=1)$ on a matrix $x$ (i.e., $\boldsymbol{Q}$ and $\boldsymbol{K}$), most elements of $x$ are non-negative, and others would be constrained to be smaller as compared to linear mappings, while improving negative value robustness and ensuring linear complexity. 
Furthermore, leveraging the proposed L2 Normalization and ELU activation for traditional dot-product attention mechanisms satisfies the aforementioned requirements:
\begin{itemize}[leftmargin=*]
\item It reduces the computational complexity of traditional dot-product attention mechanisms to $\mathcal{O}(N)$ for long-term SRSs.
\item It does not introduce additional computational costs since both the L2 Normalization and $\mathrm{elu}(\cdot,\alpha=1)$ are linear complexity operators that do not require other trainable parameters.
\item It could distinguish different items' importance in a sequence by comparing attention scores. This is because items' importance is still proportional to the corresponding element values in $\boldsymbol{B}'$.
\item It could ensure a relatively stable learning process. Since the L2 Normalization decreases the variance of attention scores by constraining the matrix's row-wise summations less than or equal to 1. Besides, the ELU function could prevent the calculation process from zero-gradient errors, thus ensuring numerical stability.
\end{itemize}
Towards this end, we formulate the final architecture of the proposed \textbf{L2 Normalized Linear Attention} (LinRec) mechanism to learn sequence representations as follows:
\begin{equation}\label{eq:LinRec}
A'(\boldsymbol{Q},\boldsymbol{K},\boldsymbol{V})=\rho_1\big(\mathrm{elu}(\boldsymbol{Q})\big)\Big(\rho_2\big(\mathrm{elu}(\boldsymbol{K})\big)^\mathrm{T} \boldsymbol{V}\Big),
\end{equation}
where the row-wise mapping is defined as $\rho_1(\boldsymbol{Q}_i)=\frac{1}{\sqrt{d}\Vert \boldsymbol{Q}_i\Vert_2}\boldsymbol{Q}_i$ for $\forall i\in[N]$, and the column-wise mapping is defined as $\rho_2(\boldsymbol{K}_j)=\frac{1}{\sqrt{N}\Vert \boldsymbol{K}_j\Vert_2}\boldsymbol{K}_j$ for $\forall j\in[d]$, where $\boldsymbol{Q}_i$ is $i$-th row of $\boldsymbol{Q}$ and $\boldsymbol{K}_j$ is $j$-th col of $\boldsymbol{K}$.
Therefore, we could integrate Eq.~(\ref{eq:LinRec}) into existing Transformer-based recommenders~\cite{liu2021augmenting,de2021transformers4rec,yang2022multi,du2022contrastive} to generate sequence representations as follows:
\begin{equation}
\begin{aligned}
    \mathrm{head}_i&=\boldsymbol{A'}(\boldsymbol{H}^l\boldsymbol{W}_Q^{(i)},\boldsymbol{H}^l\boldsymbol{W}_K^{(i)},\boldsymbol{H}^l\boldsymbol{W}_V^{(i)}),\\
    \mathrm{MH}(\boldsymbol{H}^l)&=\mathrm{Concat}(\mathrm{head}_1,\cdots,\mathrm{head}_h)\boldsymbol{W}_O,\\
    \boldsymbol{S}^{l-1}&=\mathrm{LayerNorm}(\boldsymbol{H}^{l-1}+\mathrm{Dropout}(\mathrm{MH}(\boldsymbol{H}^{l-1}))),\\
    \boldsymbol{H}^{l}&=\mathrm{LayerNorm}(\boldsymbol{S}^{l-1}+\mathrm{Dropout}(\mathrm{FNN}(\boldsymbol{S}_{l-1})))),\\ 
    \boldsymbol{H}^1&=\boldsymbol{E}
    \: ;\:
\boldsymbol{H}=\boldsymbol{H}^{L}\boldsymbol{W}_{L}+\boldsymbol{b}_{L},\\
 \end{aligned}
\end{equation}
where $\boldsymbol{W}_Q^{(i)},\boldsymbol{W}_K^{(i)},\boldsymbol{W}_V^{(i)}\in\mathbb{R}^{d\times d}$ are weight matrices at head $i$, $\boldsymbol{W}_O$ is weight matrix at multi-head (MH) block, LayerNorm refers to layer normalization function~\cite{ba2016layer}, $\boldsymbol{H}^{l}$ is hidden value at layer $l$ ($l=1,\cdots,L$) which is iteratively generated until $L$, FNN$(\cdots)$ is Feed-Forward Network, $\boldsymbol{W}_{L}\in\mathbb{R}^{hd\times d},\boldsymbol{b}_{L}\in\mathbb{R}^{d}$ are weight and bias respectively. Eventually, we get sequence representation $\boldsymbol{H}\in\mathbb{R}^{N\times d}$.

\subsection{Prediction and Model Optimization}
After obtaining item representations (i.e., $\boldsymbol{H}\in\mathbb{R}^{N\times d}$) in a sequence, we make the next-item recommendation by calculating a probability distribution of the next-item of the whole item set. At time $t$, for each candidate item $v_i$, we can calculate its recommendation score:
\begin{equation}\boldsymbol{z}_i=\boldsymbol{H}_{t}(\boldsymbol{e}^s_{i})^\mathbf{T},
\end{equation}
where $\boldsymbol{H}_{t}$ it the $t$-th item's representation in a sequence, performing the sequence representation, $\boldsymbol{e}^s_{i}\in\mathbb{R}^d$ is $v_i$'s embedding. Consequently, the recommended probability of that the next-item being $v_i$, $\hat{y}_{i}$, could be computed as follows:
\begin{equation}
\hat{y}_{i}=\frac{\exp(z_i)}{\sum_{v_j \in \mathcal{V}}\exp(z_j)}.
\end{equation} 
Therefore, we formulate the sequential recommendation task as minimizing the cross-entropy of the recommendation results $\hat{y}$ to measure the difference between the prediction and ground truth as
\begin{equation}
\mathcal{L}(y,\hat{y})=y\log(\hat{y})+(1-{y})(1-\log(\hat{y})).
\end{equation}
The training target is to obtain the best parameters of the network (including $\boldsymbol{W}_Q,\boldsymbol{W}_K,\boldsymbol{W}_V$ and other parameters outside the attention layer). We will use Adam \cite{kingma2014adam} optimizer to optimize the network, a stochastic method using moment estimation.

\subsection{In-depth Analyses}
In this section, we theoretically analyze the proposed \name in terms of its advantages and complexity.

\subsubsection{\textbf{Comparison with Other Normalization Methods}}
An intuitive way to implement Eq.~(\ref{eq:new attention}) is to leverage the Softmax function twice (i.e., row- and column-wise) to perform mapping functions $\rho_1(\cdot)$ and $\rho_2(\cdot)$ via
\begin{equation}
\rho_1(\boldsymbol{Q})=\rho(\boldsymbol{Q})\: ; \:
\rho_2(\boldsymbol{K})=\rho(\boldsymbol{K}),
\end{equation}
where $\rho(\cdot)$ means the row- and column-wise Softmax for generating $\rho_1(\boldsymbol{Q})$ and $\rho_2(\boldsymbol{K})$, respectively. The Softmax function contains an exponential operation and a weight distribution operation for a $\mathbb{R}^d$ (or $\mathbb{R}^N$) sequence(vector). For a vector $\boldsymbol{x}=(x_1,\cdots,x_n)$, we could formulate it as follows:
\begin{equation}\rho(x_i)=\frac{e^{x_i}}{\sum_{j=1}^{n}e^{x_j}}.\end{equation}
It is evident that $\rho(\cdot)$ can satisfy Condition (3). Nevertheless, a fundamental property of $\rho(\cdot)$ is that the exponential function aggregates the weight into only a few positions of the sequence, which may exaggerate importance (i.e., long-tailed attention scores) of a few items in a sequence ignoring the others, thus loss useful information from the entire sequence for long-term SRSs. In contrast to such a method, the proposed L2 Normalization could generate relatively smooth attention scores to better capture information.

\subsubsection{\textbf{Model Complexity Analysis}}
We analyze the computational complexity of the \name mechanism (i.e., Eq.~\eqref{eq:LinRec}) from 
several key components: (i) Firstly, the computational cost of $\mathrm{elu}(\cdot,\alpha=1)$ 
function is activating elements in $\boldsymbol{Q}$ and $\boldsymbol{K}$, which are two $N\times d$ matrices. Thus ELU's complexity is $\mathcal{O}(Nd)$; (ii) Secondly, $\rho_1(\cdot)$ and $\rho_2(\cdot)$ perform L2 Normalization. Therefore, their total complexity is $\mathcal{O}(Nd)$; (iii) Thirdly, the matrix multiplication operator's complexity is $\mathcal{O}(Nd^2)$. Towards this end, the time complexity of \name is $\mathcal{O}(Nd^2)$. For long-term SRSs, it could significantly reduce the time complexity of Transformer-based recommenders (we omit $d$ since $N>>d$) from $\mathcal{O}(N^2)$ to $\mathcal{O}(N)$. In conclusion, our model could achieve comparable efficiency with state-of-the-art lightweight Transformer-based sequential recommendation techniques (e.g., Linformer~\cite{wang2020linformer}, Linear Transformers~\cite{katharopoulos2020transformers}, and Big Bird~\cite{zaheer2020big}).


\begin{table}
\fontsize{8}{11}\selectfont
  \caption{Statistics of the datasets.}
  \vspace{-2mm} 
  \label{tab:commands}
  \begin{tabular}{cccccc}
    \toprule
    Datasets & \# Users & \# Items & \# Interactions & \# Sparsity \\
    \midrule
    \textbf{ML-1M} & 6,041 & 3,884 & 1,000,209 & 95.74\% \\
    \textbf{Gowalla} & 64,116 & 164,533 & 2,018,421 & 99.98\% \\
    \bottomrule
  \end{tabular}
  \label{tab:Table_1}
   \vspace{-4mm} 
\end{table}

\begin{table*}
	\fontsize{8}{11}\selectfont
	\caption{Overall performance comparison.}
        \vspace{-2mm} 
	\begin{threeparttable}
	\begin{tabular}{cccccccccccc}
		\toprule
		\toprule
		\multirow{2}{*}{Datasets}&\multirow{2}{*}{Metrics}&
		\multicolumn{2}{c}{Bert4Rec} & \multicolumn{2}{c}{CORE} & 
            \multicolumn{2}{c}{FDSA} &
		\multicolumn{2}{c}{SASRec} & \multicolumn{2}{c}{SASRecF} \cr
		\cmidrule(lr){3-12}
		& & w/o & w & w/o & w & w/o & w & w/o & w & w/o & w   \cr
		\cmidrule(lr){1-12}
		\multirow{3}{*}{ML-1M}
		&Recall@10 & 0.6975 & 0.6997 & 0.5406 & 0.6088 & 0.7166 & 0.7159 & \textbf{0.7212} & \underline{0.7209} & 0.7099 & 0.7113 \cr
		&MRR       & 0.3700 & 0.3699 & 0.2051 & 0.2792 & 0.4272 & \underline{0.4299} & 0.4282 & \textbf{0.4301} & 0.4230 & 0.4249 \cr
		&NDCG@10   & 0.4483 & 0.4488 & 0.2835 & 0.3570 & 0.4964 & \underline{0.4983} & 0.4974 & \textbf{0.4997} & 0.4921 & 0.4933 \cr
		\cmidrule(lr){1-12}
		\multirow{3}{*}{Gowalla}
		&Recall@10 & 0.8717 & 0.8739 & \underline{0.9190} & \textbf{0.9242} & 0.8981 & 0.8987 & 0.9174 & 0.9171 & 0.9077 & 0.9072 \cr
		&MRR       & 0.5886 & 0.5907 & \underline{0.6948} & \textbf{0.7031} & 0.6247 & 0.6358 & 0.6690 & 0.6723 & 0.6179 & 0.6400 \cr
		&NDCG@10   & 0.6567 & 0.6589 & \underline{0.7492} & \textbf{0.7569} & 0.6907 & 0.6995 & 0.7293 & 0.7317 & 0.6880 & 0.7048 \cr
		\bottomrule
		\bottomrule
	\end{tabular}\vspace{0cm}
        \begin{tablenotes}
        \item Results of backbone models with (w) and without (w/o) \name mechanism have been shown. All improvements are \textbf{statistically significant} (i.e., two-sided t-test with $p<0.05$) over backbone models, except Recall of SASRec. In each row, the best result is bold, while the second-best result is underlined.
        \end{tablenotes}
	\end{threeparttable}
        \label{tab:Table_2}
        \vspace{-2mm}
\end{table*}

\section{Experiments}
In this section, we aim to answer the following research questions:
\begin{itemize}[leftmargin=*]
\item \textbf{RQ1}: How does integrating \name into Transformer-based recommenders perform compared with the original ones?
\item \textbf{RQ2}: Does \name outperforms other state-of-the-art efficient Transformer variants for sequential recommendation?
\item \textbf{RQ3}: How is the scalability of the \name mechanism?
\item \textbf{RQ4}: How do different components contribute to \name?
\item \textbf{RQ5}: How is the interpretation ability of \name?
\end{itemize}

\subsection{Experimental Settings}
\subsubsection{\textbf{Datasets and Evaluation Metrics}} 
To evaluate the effectiveness of the proposed \name, we conduct experiments on two publicly available datasets: (1) \textbf{ML-1M}: it contains users' ratings (about one million) on movies, and is a popular dataset for sequential recommendation. (2) \textbf{Gowalla}: it contains users' check-ins collected by SNAP group. The statistical information is presented in Table~\ref{tab:Table_1}. Identical to the previous studies~\cite{raza2022news,afsar2022reinforcement,yuan2022multi,wu2022graph,kang2018self}, we group user interactions in chronological order on all datasets and split data by the leave-one-out strategy. For example, for an input sequence $s_i = [v^i_1, \cdots, v^i_t, \cdots, v^i_{n_i}, v^i_{n_{i}+1}, v^i_{n_{i}+2}, v^i_{n_{i}+3}]$, we use ($[v^i_1, \cdots, v^i_{n_{i}}],v^i_{n_{i}+1}$) for training, ($[v^i_1, \cdots, v^i_{n_{i}+1}],v^i_{n_{i}+2}$) for validation, and ($[v^i_1, \cdots, v^i_{n_{i}+2}],v^i_{n_{i}+3}$) for testing. Moreover, a variety of common evaluation metrics, including \textit{Recall}, \textit{Mean Reciprocal Rank} (MRR), and \textit{Normalized Discounted Cumulative Gain} (NDCG), are used for top-$k$ ($k=10$) recommendations performance evaluation.

\subsubsection{\textbf{Baselines}}
To demonstrate the effectiveness of our proposed method, we compare \name with a wide range of representative state-of-the-art Transformer-based recommenders, including traditional (i.e., $\mathcal{O}(N^2 d)$ complexity) and efficient (i.e., $\mathcal{O}(N d^2)$ complexity) Transformer-based methods. 

\noindent \textit {\textbf{Traditional Transformer}}: 
(1) \textbf{BERT4Rec}~\cite{sun2019bert4rec} uses bidirectional Transformer to fuse sequence information and a mask \& fill self-supervised task to enhance sequence representations.
(2) \textbf{CORE}~\cite{hou2022core}: leverages a simple yet efficient framework to unify representation space, thus generating better representations bridging the inconsistency between items and sessions.
(3) \textbf{FDSA}~\cite{zhang2019feature}: equips different self-attention blocks to learn item- and feature-wise representations, thus enhancing sequence representation learning.
(4) \textbf{SASRec}~\cite{kang2018self}: uses the traditional self-attention block (i.e., multi-head attention) for generating sequence representation. 
(5) \textbf{SASRecF}\footnote{https://www.recbole.io/docs/index.html}: improves the learning capabilities of SASRec by introducing item features as additional contextual information.

\noindent \textit {\textbf{Efficient Transformer}}:
(1) \textbf{Linear Transformer}~\cite{katharopoulos2020transformers}: rearranges the computational operation of the dot product in the self-attention mechanism. The authors find that the modified process contains a repetitious operation that they can reuse to reduce complexity.
(2) \textbf{Efficient Attention}~\cite{shen2021efficient}: separates the dot-product attention into two processes conducted by different scaling mappings.

\subsubsection{\textbf{Implementation Details}}

Identical to previous studies~\cite{sun2019bert4rec, kang2018self, hou2022core}, the trainable parameters are initialized with a Gaussian distribution. We optimize \name with other Transformer-based recommenders with Adam~\cite{kingma2014adam} and use the default learning rate of 0.001 and default mini-batch size of 2,048 (we decrease the mini-batch size on Gowalla to 512 to avoid GPU memory errors). Suggesting by the original papers~\cite{hou2022core, sun2019bert4rec, zhang2019feature, kang2018self, katharopoulos2020transformers, shen2021efficient}, we set the hyper-parameters for all models, including Transformer layer as $L=2$, attention head as $h=8$, and inner size (e.g., FNN layers) as $256$. In particular, we set dimension size as $d=128,64$ and maximum sequence length $N=200,100$ on ML-1M and Gowalla datasets, respectively. We further pad short-term sequences (i.e., those with $n_i<N$) by zero to ensure $n_i>d$ for long-term sequential recommendation. 
The maximum number of training epochs is $100$. 
Moreover, we adopt the early-stopping training strategy \textcolor{black}{if the NDCG@10 performance on the validation set decreases for $10$ continuous epochs.}
We implement our model\footnote{\url{https://github.com/Applied-Machine-Learning-Lab/LinRec}} in PyTorch 1.13, Python 3.7.15, and RecBole 1.1.1~\cite{zhao2021recbole}. 

\subsection{Traditional Transformer Comparison (\textbf{RQ1})}

To demonstrate the effectiveness of \name for improving recommendation performance, we integrate \name into other representative Transformer-based recommenders (e.g., BERT4Rec and SASRec). We report the main experimental results in Table~\ref{tab:Table_2}, 
where we can draw a few interesting observations as follows:
\begin{itemize}[leftmargin=*]
    \item Generally, integrating with \name, Transformer-based recommenders perform significantly better than the original ones in most cases, demonstrating the effectiveness of the proposed method. We attribute such improvements to \name could improve the long-term information capturing capabilities of Transformer by generating a relatively smooth attention matrix. 
    \item Comparing with other recommenders, integrating \name into SASRec (LinRec+SASRec) consistently yields best performance on ML-1M, while LinRec+CORE resulting the best performance on Gowalla. Such results present that CORE is more suitable for sparse datasets (i.e., short-term sequences), while equipping with the proposed \name could significantly improve the performance on both ML-1M and Gowalla datasets, again demonstrating the superiority of \name.  
\end{itemize}

\subsection{Efficient Transformer Comparison (\textbf{RQ2})}
To demonstrate the superiority of \name over state-of-the-art efficient Transformer methods, we take the SASRec as a backbone, and report the experimental results on ML-1M and Gowalla datasets in Table~\ref{tab:Table_3}. All improvements are statistically significant by performing two-sided $t$-test with $p<0.05$. Accordingly, we have a few observations as follows:
\begin{itemize}[leftmargin=*]
    \item In general, the proposed LinRec+SASRec consistently outperforms other approaches on ML-1M and Gowalla datasets. In particular, LinRec+SASRec obtains an average of about 10\% improvement over other baselines on ML-1M dataset. Such results generally demonstrate the superiority of the proposed \name.
    \item Comparing the other methods (i.e., LT+SASRec and EA+SASRec) with the original SASRec model, integrating with such methods does not achieve consistent improvements. It is because these methods do not satisfy traditional dot-product attention conditions, which harms Transformer's learning abilities. In contrast, the proposed \name satisfies the properties and generates relatively smooth attention scores for sequence representation learning, which can yield better performance.
\end{itemize}

\begin{table}[t]
\centering
	\fontsize{8}{11}\selectfont
	\caption{Performance comparison with state-of-the-art efficient Transformer-based methods.}
        \vspace{-2mm}
        \resizebox{\linewidth}{!}{
        \begin{threeparttable}
	\begin{tabular}{ccccc}
		\toprule
		\toprule
		Datasets & Model & Recall@10 & MRR & NDCG@10	\cr
		\cmidrule(lr){1-5}
		\multirow{4}{*}{ML-1M}
						& LT+SASRec    & 0.6575 & 0.3359 & 0.4124  \cr
						& EA+SASRec      & 0.6873 & 0.4094 & 0.4760	\cr
						& LinRec+SASRec & \textbf{0.7209} & \textbf{0.4301} & \textbf{0.4997}	\cr
		\cmidrule(lr){2-5}
					& Imprv.  & 7.1\%$\sim$9.6\% & 9.6\%$\sim$28.0\% & 8.8\%$\sim$21.7\%\cr
		\cmidrule(lr){1-5}
		\multirow{4}{*}{Gowalla}
						& LT+SASRec  & 0.9113 & 0.6627 & 0.7230    \cr
						& EA+SASRec  & 0.9139 & 0.6577 & 0.7198	\cr
						& LinRec+SASRec & \textbf{0.9171} & \textbf{0.6723} & \textbf{0.7317}	\cr
		\cmidrule(lr){2-5}
					& Imprv.  & 0.4\%$\sim$0.6\% & 1.4\%$\sim$2.2\% & 1.2\%$\sim$1.6\%	\cr
		\bottomrule
		\bottomrule
	\end{tabular}\vspace{0cm}
	\begin{tablenotes}
        \item All improvements are statistically significant (i.e., two-sided t-test with $p<0.05$) over baseline. In each row, the best result is bold. 
        \end{tablenotes}
	\end{threeparttable}}
        \label{tab:Table_3}
        \vspace{-4mm}
\end{table}

\subsection{Scalability Study (\textbf{RQ3})}
To investigate the model efficiency, we evaluate the computational cost, including GPU memory and Time, of the proposed \name as compared to traditional Transformer baselines.

\noindent \textbf{Efficiency of Different Backbones.}
In Table~\ref{tab:Table_4}, we observe that \name mechanism exactly reduces the GPU memory and Time cost, demonstrating the high efficiency of \name. For example, \name performs extraordinarily to improve efficiency on ML-1M, which reduces the GPU memory and Time cost to approximately one-third for most baseline models. Such a result generally conforms to the theoretical analysis of linear complexity (i.e., \name could reduce complexity from $\mathcal{O}(N^2 d)$ to $\mathcal{O}(N d^2)$). 

\begin{figure}[t]
  \centering
  \includegraphics[width=1\linewidth]{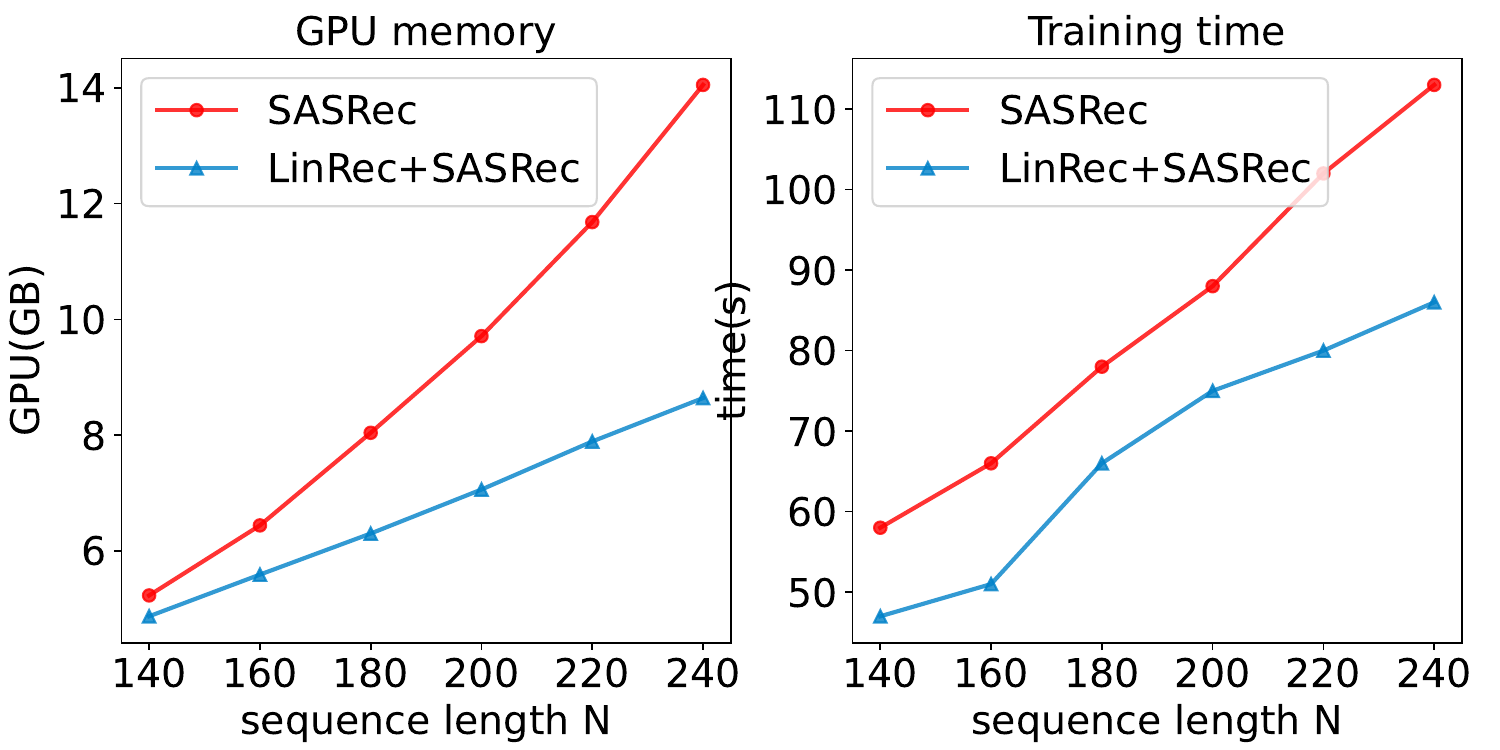}
  \vspace{-6mm}
  \caption{Parameter sensitivity.}
  \label{fig:para}
  \vspace{-3mm}
\end{figure}

\noindent\textbf{Efficiency of Different Sequence Lengths.}
To investigate the training cost of different sequence lengths for long-term SRSs, we tune maximum sequence lengths in $\{140,160,\cdots,240\}$ according to the hidden size (i.e., $d=128$, the lengths of long-term sequences typically larger than such a dimension size) settings. From Figure~\ref{fig:para}, we see the LinRec+SASRec considerably reduces the computational complexity of SASRec consistently in all cases. This is highly desirable in practice, especially for long-term sequential recommendation, where user historical interaction sequences are stored for a long time, thus learning user preferences comprehensively.

\begin{table}[t]

\centering
	\fontsize{8}{11}\selectfont  
	\caption{Efficiency comparison.}
        \resizebox{\linewidth}{!}{
	\begin{tabular}{cccccccc}
		\toprule
		\toprule
		\multirow{3}{*}{Datasets} & \multirow{3}{*}{Model} 
 & \multicolumn{2}{c}{GPU memory (GB)} & \multicolumn{4}{c}{Time cost (s/epoch)}\cr
 &  &  &  & \multicolumn{2}{c}{Training} &  \multicolumn{2}{c}{Evaluation} \cr
		\cmidrule(lr){3-8}
						& & w/o & w & w/o & w & w/o & w   \cr
		\cmidrule(lr){1-8}
		\multirow{5}{*}{ML-1M}
		& BERT4Rec    & 20.74G & 10.42G & 246s  & 157s   & 56s  & 37s\cr
		& CORE        & 19.49G & 6.79G  & 109s  & 33s    & 38s  & 16s\cr
		& FDSA        & 35.22G & 13.87G & 270s  & 99s    & 79s  & 36s\cr
		& SASRec      & 21.07G & 8.72G  & 121s  & 44s    & 41s  & 20s\cr
		& SASRecF     & 22.71G & 10.33G & 141s  & 64s    & 46s  & 23s\cr
		\cmidrule(lr){1-8}
		\multirow{5}{*}{Gowalla}						
		& BERT4Rec    & 20.10G & 19.46G & 483s  & 455s   & 516s  & 454s\cr
		& CORE        & 3.75G & 2.59G   & 126s  & 94s    & 570s  & 377s\cr
		& FDSA        & 4.81G & 3.48G   & 209s  & 116s   & 834s  & 550s\cr
		& SASRec      & 3.75G & 2.68G   & 112s  & 90s    & 434s  & 384s\cr
		& SASRecF     & 3.85G & 3.22G   & 117s  & 93s    & 445s  & 400s\cr
		\bottomrule
		\bottomrule
	\end{tabular}}
        \vspace{-1mm}
	\label{tab:Table_4}
\end{table}

\subsection{Ablation Study (\textbf{RQ4})}

\begin{figure}[t]
  \centering
  \vspace{-1mm}
  \includegraphics[width=1.0\linewidth]{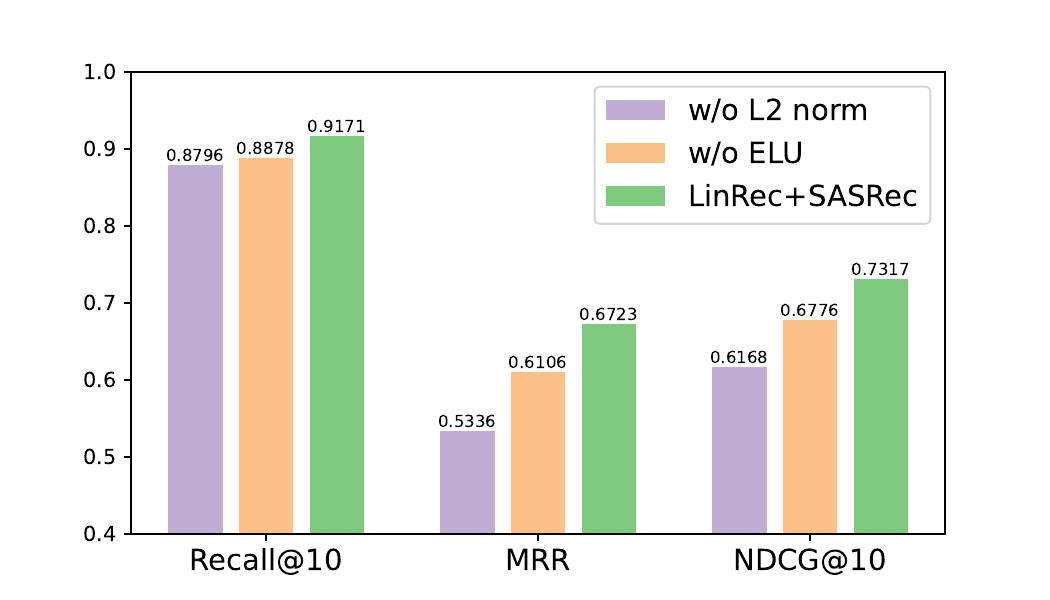}
  \vspace{-7mm}
  \caption{The results of ablation study.}
  \label{fig:ablation}
  \vspace{-4mm}
\end{figure}

To verify the contribution of each component of the proposed \name, we conduct an ablation study with two variants of the LinRec+SASRec (i.e., Eq.~(\ref{eq:LinRec})) over the Gowalla dataset, including (1) \textit{w/o L2 norm}: without the L2 normalization layer, and (2) \textit{w/o ELU}: without the ELU activation layer. Figure~\ref{fig:ablation} shows the performances of different variants in terms of Recall@10, MRR, and NDCG@10. It can be observed that each component contributes to performance. The L2 normalization is particularly crucial for \name, justifying that satisfying the normalized property (i.e., attention score summations less than or equal to 1) is the most critical part of preserving the learning capabilities of attention mechanisms. Besides, the ELU activation layer is indispensable for achieving encouraging results. Furthermore, LinRec+SASRec consistently improves performance in all cases, confirming the correctness of our design choice.  

\subsection{Case Study (\textbf{RQ5})}

\begin{figure}[t]
    \centering
    \subfigure[The long-tailed attention scores generated by SASRec, layer 1.]{
        \begin{minipage}[t]{0.475\linewidth}
            \includegraphics[width=1\linewidth]{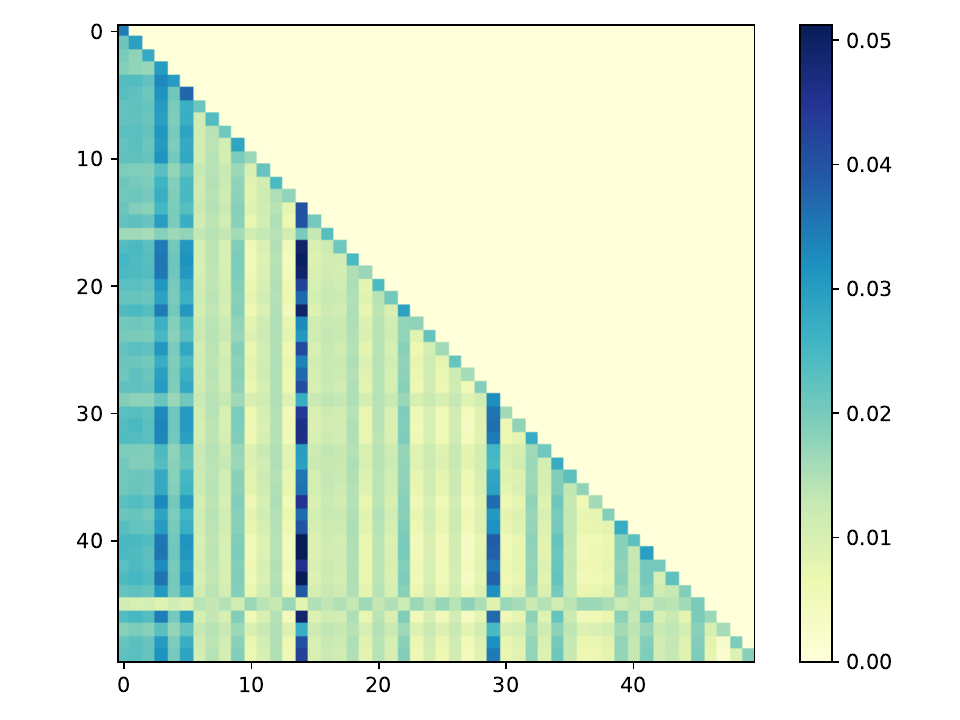}
        \label{subfig: sasrec 1}
        \end{minipage}
    }
    \subfigure[The long-tailed attention scores generated by SASRec, layer 2.]{
        \begin{minipage}[t]{0.475\linewidth}
            \includegraphics[width=1\linewidth]{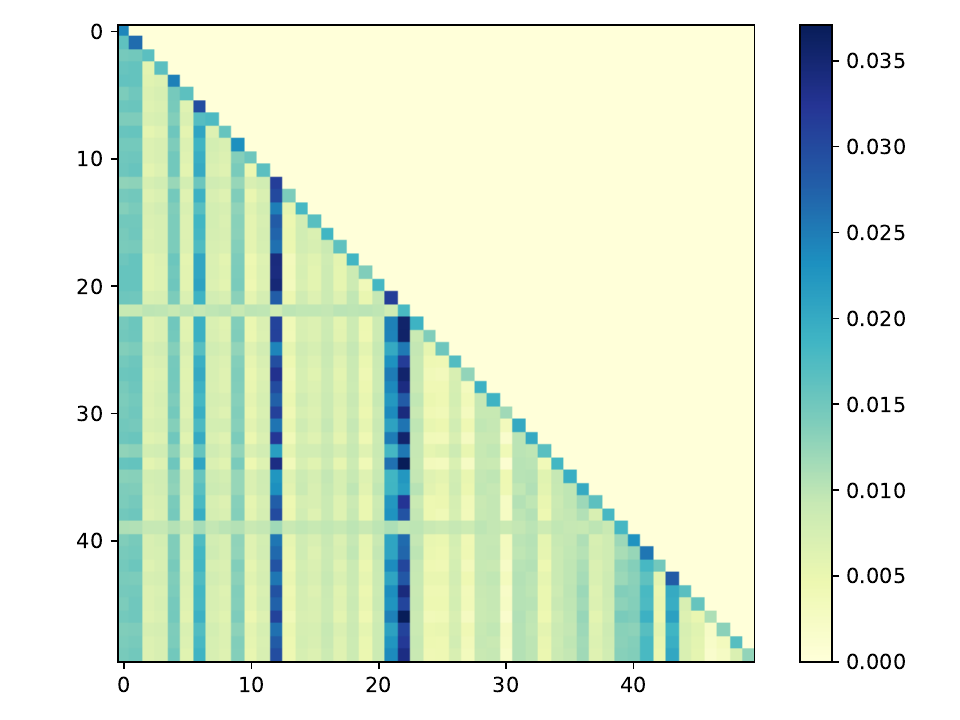}
        \label{subfig: sasrec 2}
        \end{minipage}
    }
    \subfigure[The relatively smooth attention scores generated by LinRec+SASRec, layer 1.]{
        \begin{minipage}[t]{0.475\linewidth}
            \includegraphics[width=1\linewidth]{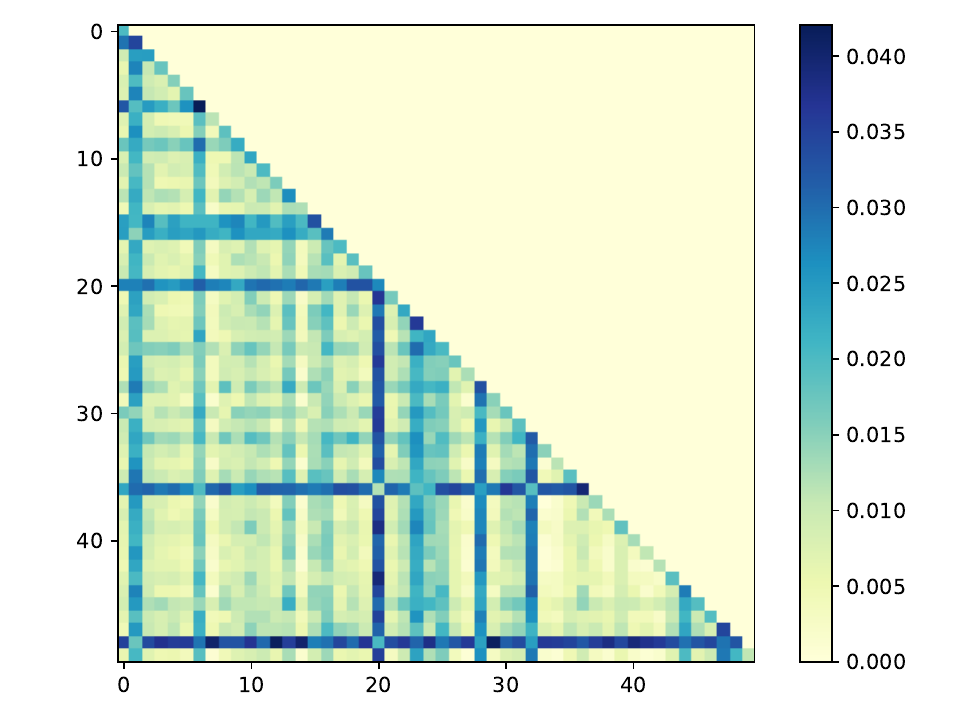}
        \label{subfig: LinRec 1}
        \end{minipage}
    }
    \subfigure[The relatively smooth attention scores generated by LinRec+SASRec, layer 2.]{
        \begin{minipage}[t]{0.475\linewidth}
            \includegraphics[width=1\linewidth]{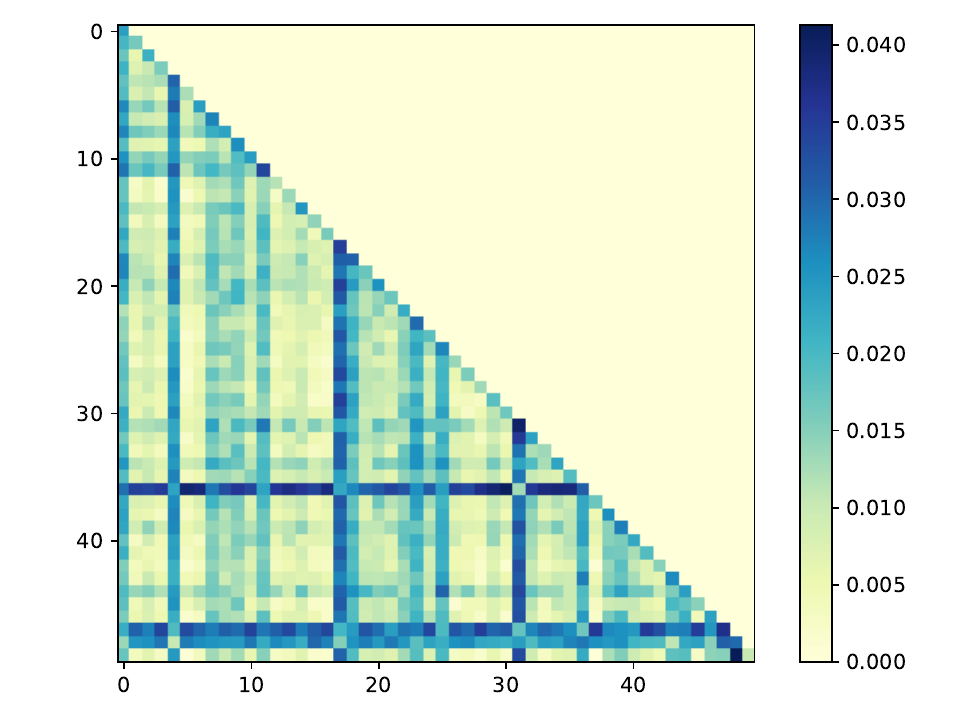}
        \label{subfig: LinRec 2}
        \end{minipage}
    }
    \vspace{-3mm}
    \caption{Heatmap of attention scores of different methods.}
    \label{fig:case study-zc}
   \vspace{-6mm}
\end{figure}

Finally, we conduct a case study to visualize the attention scores and illustrate how the attention matrix learned in \name can help to capture more information in long-term sequences. In Figure~\ref{fig:case study-zc}, we show a user whose historical interaction sequence length is 50 from the ML-1M dataset, where the darker the color is, the higher the attention score is. Moreover, we pad the sequence with zero to ensure its length is larger than the embedding dimension, and we omit the padded part for simplicity.
Comparing the heatmaps generated by SASRec and LinRec+SASRec, the proposed \name tends to generate relatively smooth attention scores, which can capture more information, including more recent or less recent items. In contrast, SASRec tends to attend to a few items, which may exaggerate such items' importance and leads to an inevitable information-losing issue for long-term sequence learning. 

Specifically, comparing with Figure~\ref{subfig: sasrec 1} and Figure~\ref{subfig: sasrec 2}, SASRec assigns larger attention scores on recent items, which is consistent with the observations in the original paper~\cite{kang2018self}, showing its limited capabilities of learning long-term sequential patterns. In contrast, comparing with Figure~\ref{subfig: LinRec 1} and Figure~\ref{subfig: LinRec 2}, \name can gradually attend to both short- and long-term patterns with increasing Transformer layers.
In conclusion, this case aligns with our motivation that \name can further improve traditional Transformer-based recommenders' learning capabilities for long-term SRSs.

\section{Related works}
In this section, we concisely review the Transformer-based SRSs and efficient Transformers to discuss the differences between the proposed \name mechanism and the related ones.
\subsection{Transformer-based SRSs}
In view of the Transformers' superior capabilities of learning sequence representations, various studies~\cite{zhang2018next,kang2018self,sun2019bert4rec,zhang2019feature,chen2019behavior,wu2020sse, hou2022core} leverage Transformers to learn sequence representation. The key idea is that Transformers can distinguish different items' importance, capturing users' long- and short-term preferences to learn better sequence representations.
Specifically, ATTRec~\cite{zhang2018next} applies attention mechanism and metric embedding to absorb users' long- and short-term information, respectively. 
SASRecc~\cite{kang2018self} captures long- and short-term sequential dynamics based on a multi-head attention mechanism.
BERT4Rec~\cite{sun2019bert4rec} is proposed to address the issue that a unidirectional model is hard to discover latent features of interaction sequences. In addition, the Cloze objective is used to improve the efficiency of the training process. 
FDSA~\cite{zhang2019feature} utilizes additional feature-level sequences in Transformer to better extract sequential patterns.
BST~\cite{chen2019behavior} uses the Transformer model to learn the sequential nature of users' behaviors.
SSE-PT~\cite{wu2020sse} applies a personalized Transformer into self-attentive
neural network architectures to improve ranking performance.
CORE~\cite{hou2022core} utilizes a representation-consistent encoder and a robust distance measuring method to improve traditional sequential encoders' (e.g., self-attention mechanism) capabilities of learning representations.

While these methods are effective, they often employ traditional dot-product attention mechanisms, which result in an inefficient learning process, particularly when calculating attention matrices.

\vspace{-1mm}
\subsection{Efficient Transformers}
To tackle the high complexity issue of Transformers, a new line of research has started to analyze and propose various efficient methods to reduce the inherently computational complexity.
Such methods mainly aim to leverage approximated algorithms to simplify the attention matrix computing process (e.g., sparse matrix computing) in self-attention mechanisms~\cite{tay2022efficient}.

The earliest research line of efficient Transformers, named fixed patterns~\cite{dai2020funnel,ainslie2020etc,child2019generating,qiu2019blockwise, beltagy2020longformer,liu2021swin}, designs attention matrices' sparse architecture based on specific patterns such as blocks and strides. 
The block-wise models~\cite{qiu2019blockwise, parmar2018image}, one approach of fixed patterns, separate the attention matrix into several blocks and use a random masking matrix to decide whether to retain each block. Another famous method of fixed patterns concerns striding approaches~\cite{child2019generating, beltagy2020longformer}, where several strides separate the masking matrix in equal intervals. 
Moreover, the Axial transformers~\cite{ho2019axial} provides a new view of applying attention mechanisms for multidimensional tensors. 
Unlike fixed patterns, learnable patterns~\cite{wang2020cluster,kitaev2020reformer,roy2021efficient,tay2020sparse,vyas2020fast} consider the masking matrices can be learned. 
Cluster-Former~\cite{wang2020cluster} uses the clustering method to learn the classes of positions in the sequence and separate the sequence into chunks. 
Then they perform the transformers in each short chunk to obtain efficiency.
Reformers~\cite{kitaev2020reformer} improve the efficiency by utilizing the LSH attention, where they use locality-sensitive hashing at each round. 

Neural memory~\cite{liu2021swin,lee2019set,beltagy2020longformer,ryoo2021tokenlearner,tay2021charformer,jaegle2021perceiver,rae2019compressive} method tries to utilize information about multiple positions in the sequence and their relevant connections. Set Transformer~\cite{lee2019set} detects and utilizes the interactions of positions in the sequence. They apply the inducing points to derive their particular attention mechanism. 
Longformer~\cite{beltagy2020longformer} uses the sliding window to capture the local context, meanwhile introducing global attention to aggregate valuable information from selected positions.
Another recently famous approach of efficient Transformers introduces low-rank approximations to reduce the complexity and simplify models~\cite{winata2020lightweight,wang2020linformer,katharopoulos2020transformers,zhu2021long,shen2021efficient,choromanski2020rethinking,peng2021random,xiong2021nystromformer,hua2022transformer}. Based on the spectrum results and theoretical analysis that self-attention is a low-rank matrix, the Linformer~\cite{wang2020linformer} derives a brief attention mechanism for reducing complexity. 
They use linear projections matrix to reduce the dimension of Key and Value matrices. 
The Linear transformers~\cite{katharopoulos2020transformers} introduce a kernel-like method. Specifically, the authors disassemble and reassemble the dot-product operation of attention mechanisms. Then they find a part of the operation repeated in each round that can be reused for efficient computing. 
Besides, some studies further explore leveraging efficient model retraining~\cite{zhang2020retrain} or sequence chunking~\cite{wang2022rete} approaches to enhance computational efficiency for long-term SRSs. They mainly focus on transferring knowledge or editing data-level structures, which do not undermine our technical contributions of reducing the computational complexity of traditional Transformer-based SRSs (e.g., dot-product attention mechanisms) when calculating attention matrices.

Despite existing efficient Transformer methods decreasing the computational complexity, they either need to sacrifice more accuracy or perform low efficiency. In contrast, the proposed \name effectively preserves attention mechanisms' advantageous properties and reduces computational costs without jeopardizing performance.

\section{Conclusion}
In this paper, we studied the problem of long-term sequential recommendation from a new perspective--how to reduce the computational complexity of traditional Transformer-based SRSs models raised by the dot-product operations. We theoretically analyze the core properties of the dot-product attention mechanism for distinguishing items' importance in sequences. Accordingly, we propose a novel \name mechanism possessing linear complexity $\mathcal{O}(N)$ while capturing more information from long-term sequences, thus generating better sequence representations for making recommendations efficiently and effectively.
Comprehensive experiments demonstrate that \name has the excellent capability of improving efficiency while keeping accuracy. Significantly, \name outperforms two state-of-the-art efficient Transformer-based methods. In addition, we provide adequate theories and discussions to support our \name mechanism, where two highlights are the equivalence conditions and the statistical interpretation. The equivalence conditions provide theoretical insights for our design choice and can assist us in generating other efficient attention mechanisms for tasks other than long-term SRSs. The Statistical Interpretation that all elements have statistical meaning provides a solid foundation for the proposed \name mechanism.


\appendix
\section{Statistical Interpretation}
\label{appendix}

Statistically, Condition (1) means the probabilities of all positions in a row add up to no more than $1$, while Condition (2) means the probabilities of all positions are more than or equal to $0$. Therefore, similar to dot-product attention, the attention scores of \name satisfy the corresponding probability properties. Considering each position $i$
in a sequence, we denote $i$'s attention to position $k$ as $\mathcal{A}_{ik}$, and the corresponding independent probability as $\mathrm{P_r}(\mathcal{A}_{ik})$. 
Then Conditions (1) and (2) can be rewritten in probability format:
\begin{equation}\sum\nolimits_{k=1}^N\mathrm{P_r}(\mathcal{A}_{ik})\le 1
\:;\:
\mathrm{P_r}(\mathcal{A}_{ik})\ge 0,\quad\forall k.
\end{equation}
Then we instead consider separating attention into $d$ sub-events by $d$ latent features, e.g., $\mathcal{B}_{i1},\cdots,\mathcal{B}_{id}$, which respect to the hidden states of attention mechanism. And we assume that the sub-events $\mathcal{B}_{ij}$
are independent. In this way, we can define the probability $\mathrm{P_r}(\mathcal{B}_{ij})$ and the conditional probability $\mathrm{P_r}(\mathcal{A}_{ik}\vert \mathcal{B}_{ij})$.
Therefore, we could rewrite the probability applying Bayes theorem as follows
\begin{equation}
\mathrm{P_r}(\mathcal{A}_{ik})=\sum\nolimits_{j=1}^d \mathrm{P_r}(\mathcal{B}_{ij})\mathrm{P_r}(\mathcal{A}_{ik}\vert \mathcal{B}_{ij}).
\end{equation}
We then substitute probabilities by elements of $\boldsymbol{Q}^{\rho}$ and $\boldsymbol{K}^{\rho}$ as 
\begin{equation}
\label{eq:conditional}
\begin{aligned}
    \mathrm{P_r}(\mathcal{B}_{ij})=\boldsymbol{Q}^{\rho}_{ij}
\: ; \:
\mathrm{P_r}(\mathcal{A}_{ik}\vert \mathcal{B}_{ij})=\boldsymbol{K}^{\rho}_{kj},\\
\mathrm{P_r}(\mathcal{A}_{ik})=\sum\nolimits_{j=1}^d \boldsymbol{Q}^{\rho}_{ij}\boldsymbol{K}^{\rho}_{kj}=\boldsymbol{B}'_{ik},
\end{aligned}
\end{equation}
where $\boldsymbol{B}'_{ik}$ (in Eq. (\ref{eq:new attention})) is the attention score of position $i$ to position $k$. Also, we can rewrite Condition (3) as
\begin{equation}\sum\nolimits_{j=1}^d\mathrm{P_r}(\mathcal{B}_{ij})\le 1
\: ; \:
\sum\nolimits_{k=1}^N\mathrm{P_r}(\mathcal{A}_{ik}\vert \mathcal{B}_{ij})\le 1.
\end{equation}
Thus, all components of \name have statistical meaning. 
In addition, we observe that Eq. \eqref{eq:conditional} holds for any $i$. Then we reveal an interesting property of the \name mechanism:
\begin{equation}
\mathrm{P_r}(\mathcal{A}_{i_1 k}\vert \mathcal{B}_{i_1 j})=\mathrm{P_r}(\mathcal{A}_{i_2 k}\vert \mathcal{B}_{i_2 j})=\boldsymbol{K}^{\rho}_{kj}.
\end{equation}
Accordingly, such a shared conditional probability can significantly reduce parameter numbers, resulting in high-efficiency computing.

\begin{acks}
This work was supported by Ant Group Research Fund. It was also partially supported by APRC - CityU New Research Initiatives (No.9610565, Start-up Grant for New Faculty of City University of Hong Kong), SIRG - CityU Strategic Interdisciplinary Research Grant (No.7020046, No.7020074), CityU-HKIDS Early Career Research Grant (No.9360163), InnoHK initiative, The Government of the HKSAR, and Laboratory for AI-Powered Financial Technologies.
\end{acks}

\clearpage
\bibliographystyle{ACM-Reference-Format}
\bibliography{sample-base}

\end{document}